\documentclass[prb,twocolumn,showpacs,preprintnumbers,amsmath,amssymb,citeautoscript]{revtex4-1}
\usepackage{amssymb,amsfonts,amsmath}
\usepackage{graphicx}
\usepackage{bm}
\usepackage{float}

\begin{document}

\title{Cyclotron resonance study of quasiparticle mass and scattering rate in the hidden-order and superconducting phases of URu$_2$Si$_2$}

\author{S. Tonegawa$^{1}$, K. Hashimoto$^{1,*}$, K. Ikada$^{1}$, Y. Tsuruhara$^{1}$, Y.-H. Lin$^{1}$, H. Shishido$^{1,\dag}$, Y. Haga$^{2}$, \\
T.\,D. Matsuda$^{2,\ddag}$, E. Yamamoto$^{2}$, Y. Onuki$^{2,3}$, H. Ikeda$^{1}$, Y. Matsuda$^{1}$, and T. Shibauchi$^{1}$}
\affiliation{
$^1$Department of Physics, Kyoto University, Kyoto 606-8502, Japan\\
$^2$Advanced Science Research Center, Japan Atomic Energy Agency, Tokai 319-1195, Japan\\
$^3$Faculty of Science, University of the Ryukyus, Nishihara, Okinawa 903-0213, Japan
}
\date{\today}

\begin{abstract}

The observation of cyclotron resonance in ultra-clean crystals of URu$_2$Si$_2$ [S. Tonegawa {\it et al.} Phys. Rev. Lett. {\bf 109}, 036401 (2012)] provides another route besides quantum oscillations to the determination of the bulk electronic structure in the hidden order phase. We report detailed analyses of the resonance lines, which fully resolve the cyclotron mass structure of the main Fermi surface sheets. A particular focus is given to the anomalous splitting of the sharpest resonance line near the [110] direction under in-plane magnetic-field rotation, which implies peculiar electronic structure in the hidden order phase. The results under the field rotation from [110] toward [001] direction reveal that the splitting is a robust feature against field tilting from the basal plane. This is in sharp contrast to the reported frequency branch $\alpha$ in the quantum oscillation experiments showing a three-fold splitting that disappears by a small field tilt, which can be explained by the magnetic breakdown between the large hole sphere and small electron pockets. Our analysis of the cyclotron resonance profiles reveals that the heavier branch of the split line has a larger scattering rate, providing evidence for the existence of hot-spot regions along the [110] direction. These results are consistent with the broken fourfold rotational symmetry in the hidden-order phase, which can modify the interband scattering in an asymmetric manner. We also extend our measurements down to 0.7\,K, which results in the observation of cyclotron resonance in the superconducting state, where novel effects of vortex dynamics may enter. We find that the cyclotron mass undergoes no change in the superconducting state. In contrast, the quasiparticle scattering rate shows a rapid decrease below the vortex-lattice melting transition temperature, which supports the formation of quasiparticle Bloch state in the vortex lattice phase. 
\end{abstract}

\maketitle

\section{Introduction}
The heavy-fermion compound URu$_2$Si$_2$ has attracted much attention for its mysterious phase, the so called hidden-order (HO) phase, whose order parameter is not yet identified despite intense experimental and theoretical efforts for more than a quarter century \cite{Myd11}. The HO transition at $T_{\rm HO}=17.5$\,K \cite{Pal85,Map86,Sch86} accompanies a huge amount of entropy loss, but no magnetic ordering accounting for this change has been observed \cite{Bro87,Mat01,Tak07}. Without knowing which symmetry is broken in the ordered phase, many theoretical proposals for the HO parameter have been made \cite{Myd11,Kis05,Var06,Hau09,Cri09,Har10,Dub11}. 
Recent magnetic torque measurements reveal the in-plane anisotropy of magnetic susceptibility \cite{Oka11,Shi12}, which suggests that below the HO phase transition the fourfold rotational symmetry in the tetragonal URu$_2$Si$_2$ is broken. This newly suggested rotational symmetry breaking has raised several theoretical proposals \cite{Tha11,Pep11,Fuj11,Opp11,Ike12,Rau12,Han12,Ris12,Das12,Cha13}, and thus calls for further experimental verifications by using other techniques.

The nature of electronic orders in metals and semiconductors is, in general, closely related to the electronic structure, and the most essential information is the structure of Fermi surface (FS). In the case of URu$_2$Si$_2$, the large loss of entropy \cite{Pal85,Map86,Sch86} below $T_{\rm HO}$ signifies that a large portion of the FS is gapped in the hidden-order phase, which has also been supported by the transport \cite{Sch87,Beh05,Kas07} and tunneling \cite{Sch10,Ayn10} measurements.  For the understanding of the nature of hidden order, it is indispensable to determine how the electronic structure changes with the gap formation. 
In addition to this electronic excitation gap, neutron inelastic scattering experiments\cite{Wie07} revealed that the gap is formed in the magnetic excitations characterized by two wave vectors: $\bm{Q}_{C} = (1,0,0)=(0,0,1)$ with an energy gap of $E_0\cong 1.9$\,meV and $\bm{Q}_{IC} = (1\pm 0.4,0,0)$ with an energy gap of $E_1\cong 4-5.7$\,meV. 

Quantum oscillation experiments \cite{Ohk99,Has10,Aok12,Jo07,Shi09} that can yield direct information on the FS structure have revealed the existence of small pockets in the hidden order phase. However, the total density of states of these pockets is significantly smaller than the estimate from the electronic specific heat, which indicates there must be some FS sheets with heavy mass missing in these experiments. The most recent measurements of Shubnikov-de Haas (SdH) effect \cite{Has10,Aok12} have suggested that the FS is quite similar to that in the antiferromagnetic phase, which is known to be induced by applying pressure \cite{Ami07}. In this pressure-induced antiferromagnetic phase, large staggered moment along the $c$ axis has been observed with the wave vector of $\bm{Q}_{C}$, indicating the zone folding associated with the lattice doubling. Such zone folding that has also been suggested by the recent angle-resolved photoemission study in the hidden-order phase \cite{Yos10,Men13} when compared with the FS above $T_{\rm HO}$ \cite{Kaw11,Bia09,San09}, which further support the similarity between the antiferromagnetic and hidden-order states. However, the hidden order phase, in which no magnetic ordering has been detected, is separated from the pressure-induced antiferromagnetic phase by a phase transition. Then the key question to address is what is the peculiar signature in the electronic structure of the hidden order phase, and in particular it is important to clarify how this is related to the rotational symmetry breaking suggested by the torque measurements. 

Cyclotron resonance (CR) is another powerful probe of the detailed FS structure. The CR stems from the transition between Landau levels formed by the quantized cyclotron motion of the conduction electrons. It gives direct information on the effective cyclotron mass $m^*_{\rm CR}$ of electrons moving along extremal orbits on FS sheets through the simple relation $m^*_{\rm CR}=eH_{\rm CR}/\omega$, where $\omega=2\pi f$ is the microwave angular frequency and $H_{\rm CR}$ is the resonance field. The cyclotron mass $m^*_{\rm CR}$ may be different from the thermodynamic mass or the mass $m^*_{\rm QO}$ deduced from quantum oscillation measurements. In particular, in one-component translationally invariant systems $m^*_{\rm CR}$ is not renormalized by the electron-electron interaction (the Kohn's theorem \cite{Koh61}) due to the cancellation by the backflow effect, so in this case $m^*_{\rm CR}$ can be directly compared with the band mass which can be calculated without considering electron correlations. However, in solids this theorem can be violated \cite{Kan97} especially for the heavy-fermion systems with interacting conduction and $f$ electrons \cite{Var86} and for multiband systems \cite{Yos05,Kim11}. Therefore the momentum dependence of $m^*_{\rm CR}$ in each FS should contain important information on the electron correlations in these systems. 

Recent progress of the high-quality single crystal growth of URu$_2$Si$_2$ has lead to the first observation of CR among heavy-fermion materials\cite{Tonegawa2012}. In this paper, we describe the detailed analysis on this observation, which reveals the full determination of the main FS sheets including the missing heavy band. The in-plane angle dependence of $m^*_{\rm CR}$ shows an unexpected splitting for the sharpest resonance line, which has been assigned to the hole FS pocket $\alpha$ with the largest volume and mobility. This anomalous two-peak splitting is found near the [110] direction and the two-peak structure survives against field inclination toward [001] direction in the measured angle range up to 30$^\circ$. We compare our CR results with the quantum oscillation results, from which we estimate the electronic specific heat coefficient that can account for more than 80\% of the experimental value. The SdH measurements also report the splitting of the $\alpha$ branch for the oscillation frequency which is a measure of the size of the FS, but we propose that this originates from the magnetic breakdown and that our CR mass split has a different origin. The CR profile analysis allows us to estimate the scattering rates of each orbit, which reveals the emergence of hot spots with larger scattering and heavier mass along the [110] direction. This electronic structure anomaly gives strong support for the rotational symmetry breaking in the HO phase, providing a stringent constraint on the symmetry of the hidden order. 

Another important aspect of URu$_2$Si$_2$ is that the HO phase hosts the unconventional superconducting (SC) phase below the transition temperature $T_{\rm SC} =1.4$\,K at ambient pressure. The cyclotron mass $m^*_{\rm CR}$ in the superconducting state has also been a subject of theoretical debate \cite{Drew1995,Kopnin2001}. By considering the ac dynamics of superconducting vortices, a theory predicts the violation of the Kohn's theorem and a peculiar temperature dependence of the resonance frequency in clean type-II superconductors \cite{Kopnin2001}. Experimentally, however, this point has not yet studied mainly because the observation of CR in the superconducting state is difficult due to the limitation of microwave penetration depth which is usually short. By using a $^3$He microwave cavity we are able to observe CR in the SC phase of URu$_2$Si$_2$. Contrary to the proposed temperature dependence, we find that the mass does not show any significant change below $T_c$. We rather find that the scattering rate at low temperatures exhibits characteristic temperature dependence; it shows non-Fermi liquid-like quasi $T$-linear dependence followed by a sudden decrease below the vortex-lattice melting transition temperature, which has been determined by the resistivity measurements\cite{Oka08}. This supports the formation of a coherent quasiparticle Bloch state in the vortex lattice phase.


\section{methods}

\subsection{Microwave measurements}
High-quality single crystals of URu$_2$Si$_2$ used in this study were grown by the Czochralski pulling method and applying the solid state electro-transport method under ultra-high vacuum \cite{Mat08,Mat11}. The very high residual-resistivity-ratio ($\sim700$) \cite{Kas07,Oka08} indicates that the impurity scattering rate is very low which is important for observation of CR.  

The CR experiments were carried out by using three different cylindrical Cu cavities \cite{Shibauchi94,Shibauchi97}, whose resonance frequencies for the TE$_{011}$ mode are 28, 45, and 60\,GHz. For the 28-GHz cavity, we also use the TE$_{012}$ mode at 38\,GHz. The quality factors of the cavities are $2-4\times10^4$ without the sample. The plate-like crystal with dimensions of $2.1\times0.58\times0.10$\,mm$^3$ is placed at an antinode position of the microwave magnetic field $H_\omega$. The 28 and 45-GHz cavities have sapphire hot-fingers \cite{Shibauchi94}, which enables us to control the temperature of the sample with keeping the cavity temperature at the liquid $^4$He temperature. The 45-GHz cavity has a $^3$He pot system which can cool the sample to $\sim \ $0.7 K. For the 60-GHz cavity measurements, the sample is placed by a sapphire rod attached to the cavity and the cavity temperature is varied. The frequency response of the cavity is measured by a scalar network analyzer, and the microwave loss $1/Q$ and resonant frequency $f$ are recorded as a function of dc field $H$ with and without the sample. The dc magnetic field $\bm{H}$ is applied perpendicular to the alternating currents $\bm{J}_\omega$ induced by microwave [Fig.\:\ref{Fig:HO}(a)]. In a classical picture, the electrons undergo cyclotron motion with velocity perpendicular to $\bm{H}$ and are accelerated by the microwave. The field-angle is controlled by manually rotating the sample at room temperature. 

The changes in $\Delta 1/Q$ and $\Delta f$ as a function of dc field $H$ show the multiple cyclotron resonances [Fig.\:\ref{Fig:HO}(b)]. The CR occurs near the surface within the microwave skin depth $\delta=(2\rho/\mu_0\omega)^{1/2}$ ($\sim 0.3\,\mu$m for 28\,GHz at 1.7\,K where the dc resistivity $\rho$ is $\sim 1\,\mu\Omega$cm in our crystal) when the frequency of the cyclotron motion coincides with the microwave frequency. We note that unlike conventional CR in metals \cite{Azb58}, heavy mass and small carriers (namely, slow Fermi velocity $v_F$) in URu$_2$Si$_2$ result in that the cyclotron radius $r_c=v_F/\omega$ 
may become shorter than $\delta$ in our measurement frequency range.

\subsection{Band-structure calculations}
Electronic band-structure is calculated in two steps. First, the {\it ab initio} calculations are performed for the paramagnetic state of URu$_2$Si$_2$ by using the \textsc{Wien2k} package \cite{Wien2k}, in which the relativistic full-potential (linearized) augmented plane-wave (FLAPW) $+$ local orbitals method is implemented. The crystallographical parameters are the space group No.139, $I4/mmm$, the lattice constants, $a = 4.126$\,\AA, $c = 9.568$\,\AA, and Si internal position, $z = 0.371$ \cite{Cor85}. Then the Fermi surface in the antiferromagnetic state was obtained by applying several values of effective field, and the Brillouin zone is folded to the space group No.123, $P4/mmm$. The obtained Fermi surface is essentially consistent with the previous density functional band-structure calculations \cite{Opp10}. 

\section{cyclotron resonance in the hidden order phase}

\subsection{Determination of the cyclotron mass}

\begin{figure}[t]
\includegraphics[width=1.0\linewidth]{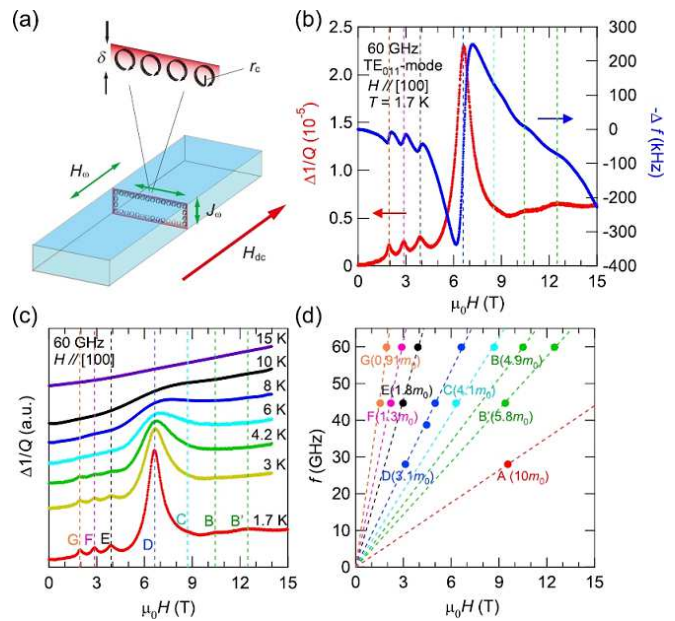}
\caption{(Color online) Observation of cyclotron resonance in the hidden-order phase of URu$_2$Si$_2$. (a) Schematic configuration of the microwave measurements. The crystal is placed inside the cavity, where the microwave field component $H_{\omega}$ has an antinode for the TE$_{011}$ and TE$_{012}$ modes. The dc field $H$ is applied parallel to $H_\omega$, which excites the microwave current $J_\omega$ near the sample surface in the region characterized by the skin depth $\delta$ (red shades). When the frequency of the cyclotron motion with a radius $r_c$ (black loops) coincides with the microwave frequency $f$, the cyclotron resonance occurs. (b) Magnetic-field dependence of the change in the microwave dissipation $\Delta 1/Q$ (red, left axis) and the frequency shift $\Delta f$ (blue, right axis) of the 60-GHz cavity resonator containing a single crystal at $1.7$\,K. The dc field is along the $[100]$ direction. The weak field dependence measured without the crystal has been subtracted. The dotted lines mark the resonance fields. (c) Field dependence of $\Delta 1/Q$ at several different temperatures. Each curve is shifted vertically for clarity. (d) Relation between the measured frequencies and the resonance fields for $\bm{H} \parallel [100]$. }
\label{Fig:HO}
\end{figure}

\begin{figure*}[t]
\includegraphics[width=0.75\linewidth]{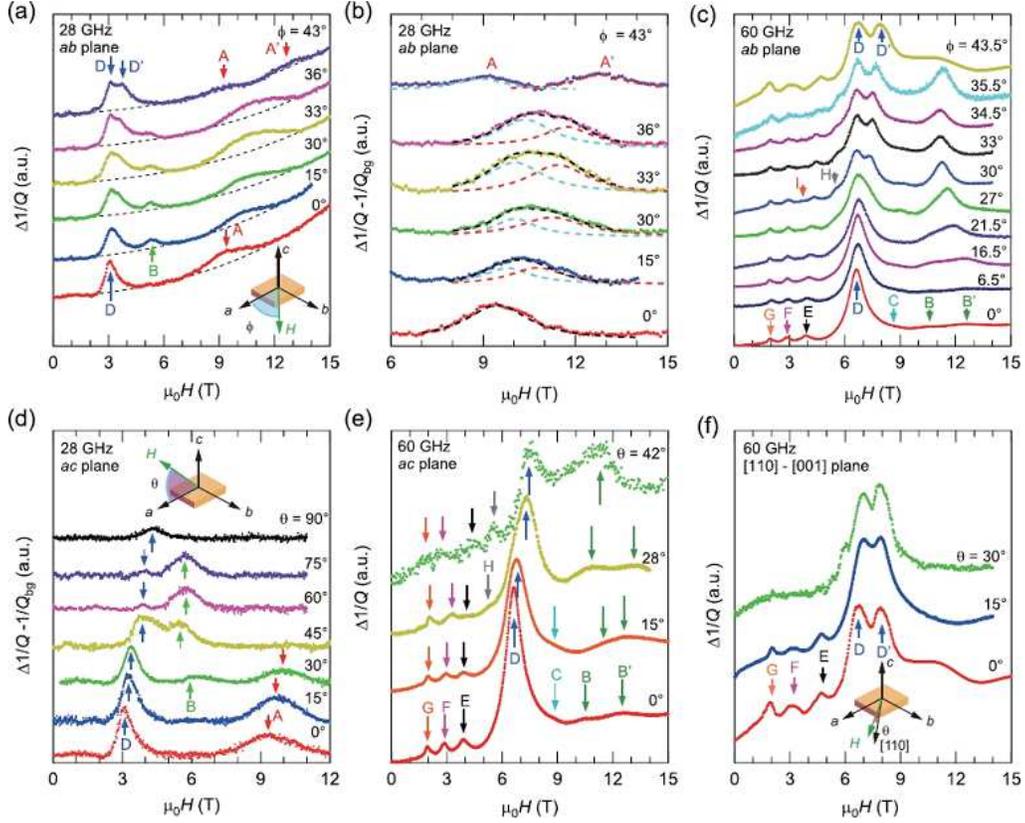}
\caption{(Color online) Cyclotron resonance in URu$_2$Si$_2$ for $\bm{H}$ applied along several different directions. 
(a) $\Delta 1/Q$ under in-plane field rotation at 28\,GHz. Here $\phi$ is the field angle from the $[100]$ direction in the $ab$ plane. (b) Smooth polynomial background field dependence $1/Q_{\rm{bg}}$ (dotted lines in (a)) has been subtracted. At finite angles, the line A shows a broadened shape, which can be fitted to two Lorentzian functions (dashed lines). 
(c) $\Delta 1/Q(H)$ at 60\,GHz for several angles $\phi$ in the $ab$ plane. (d) 60\,GHz data in the $ac$ plane. Here $\theta$ is the field angle from the $ab$ plane. (e) 28\,GHz data in the $ac$ plane. (f) Similar data at 60\,GHz but in the $[110]$-$[001]$ plane.}
\label{Fig:HO_CR}
\end{figure*}

Figure\:\ref{Fig:HO}(b) shows the microwave data for $\bm{H}//[100]$ at 1.7\,K representing the observation of the CR. The microwave power dissipation $\Delta1/Q$ as a function of applied dc field shows several peaks and at the same fields the frequency shift $\Delta f$ shows rapid changes, which are expected from the Kramers-Kronig relations between real and imaginary parts of the response functions. These results clearly indicate that the multiple resonances occur in this field range. These resonances show rapid broadening with increasing temperature [Fig.\:\ref{Fig:HO}(c)], which rules out the electron paramagnetic resonance as the origin of anomalies. Measurements by using different cavities or different modes [Fig.\:\ref{Fig:HO}(d)] clearly demonstrate that the resonance fields are proportional to the measurement frequency. All of these features establish that these anomalies are due to the CR. The observed seven CR lines are labelled as A to G in the order of corresponding $m^*_{\rm CR}$ from the heaviest [Figs.\:\ref{Fig:HO}(d) and \ref{Fig:HO_CR}]. 

The cyclotron resonance data set measured at 28 and 60\,GHz in the $ab$, $[110]$-$[001]$, and $ac$ planes are shown in Figs.\:\ref{Fig:HO_CR}(a)-(f). Here the sample temperature is $\sim 1.7$\,K. The cyclotron resonance occurs when the relation $\omega=\omega_c (=eH_{CR}/m^*_{\rm CR})$ is satisfied, from which the angle dependence of $m^*_{\rm CR}$ can be extracted as shown in Fig.\:\ref{Fig:HO_mass}. For the condition $\omega_c\tau>1$ (where $\omega_c$ is the cyclotron angular frequency and $\tau$ is the scattering time) the microwave dissipation $\Delta 1/Q(H)$ has a peak at $H_{\rm CR}$, which is described as follows. 

The dissipation is proportional to the real part of the complex conductivity $\sigma(\omega)$, which is given by the Drude model in a simple metal with the carrier number $n$ and effective mass $m^*$ \cite{Dresselhaus55}:
\begin{equation}
\sigma(\omega) = \sigma_0\frac{1+\mathrm{i}\omega\tau}{(1+\mathrm{i}\omega\tau)^2+(\omega_c\tau)^2}, \qquad
\sigma_0=\frac{ne^2\tau}{m^*}.
\end{equation}
The real part is then given by
\begin{equation}
\mathrm{Re}\{\sigma(\omega)\} = \frac{\sigma_0}{2}\left[\frac{1}{(\omega-\omega_c)^2\tau^2+1}+\frac{1}{(\omega+\omega_c)^2\tau^2+1}\right].
\end{equation}
This is the sum of two Lorentzian functions which have two peaks at $\omega=\pm\omega_c$ with the width determined by $1/\tau$. When the resonance peak is sharp enough ($\omega_c\tau\gg1$), the contribution from the $\omega=-\omega_c$ peak becomes negligible near the actual resonance at $\omega=\omega_c$, and then the peak in $\Delta 1/Q(H)$ can be approximated by the simple Lorentzian 
\begin{equation}
\Delta 1/Q(H) \propto \frac{1}{(H-H_{CR})^2+(\Delta H/2)^2},
\end{equation}
where the normalized full width at half maximum (FWHM) $\Delta H/H_{CR}$ is given by $2/\omega_c\tau$.

From the normalized FWHM we estimate that $\omega_c\tau$ reaches $\sim 20$ at low temperatures for the sharpest line D. In addition to the resonance peaks, the field dependent surface resistance contributes to $\Delta 1/Q(H)$ as well. Since URu$_2$Si$_2$ is a compensated metal with equal volumes of electron and hole carries, the magnetoresistance is large at low temperatures for high-quality crystals with large $\tau$ \cite{Kas07}. This gives noticeable smooth background signals, as evident especially for 28\,GHz. To resolve the cyclotron resonance lines at high fields, we therefore subtract this background field dependence by using polynomial functions [dashed lines in Fig.\:\ref{Fig:HO_CR}(a)]. For the resonance line A, we fit the subtracted data by the two Lorentzian functions with different $H_{\rm CR}$ [Fig.\:\ref{Fig:HO_CR}(b)].

\subsection{Angle dependence of the cyclotron mass}

\begin{figure}[t]
\includegraphics[width=\linewidth]{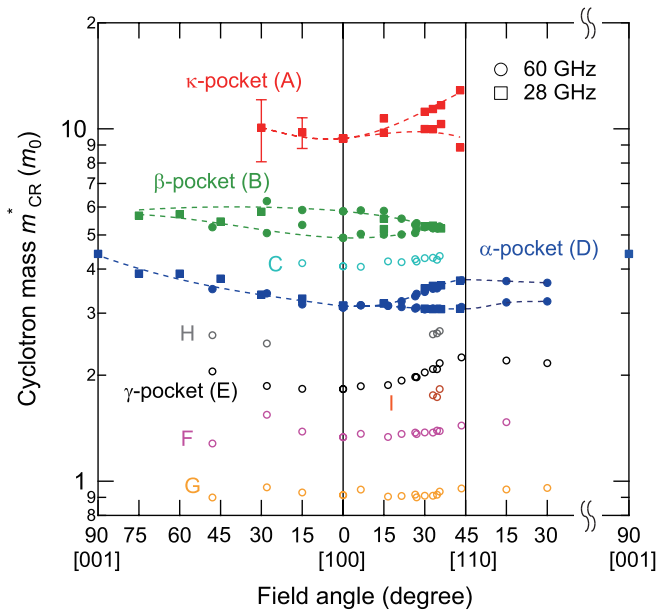}
\caption{(Color online) Structure of the cyclotron masses $m^*_{\rm CR}$ for each Fermi surface sheet in URu$_2$Si$_2$. Cyclotron masses as a function of field angle for the $ac$-plane (left), $ab$-plane (center) and $[110]-[001]$ plane (right) rotations. The solid (open) symbols are for the resonance lines with large (small) intensities. The dashed lines are guides to the eyes for the main three bands.}
\label{Fig:HO_mass}
\end{figure} 

\begin{figure*}[t]
\includegraphics[width=\linewidth]{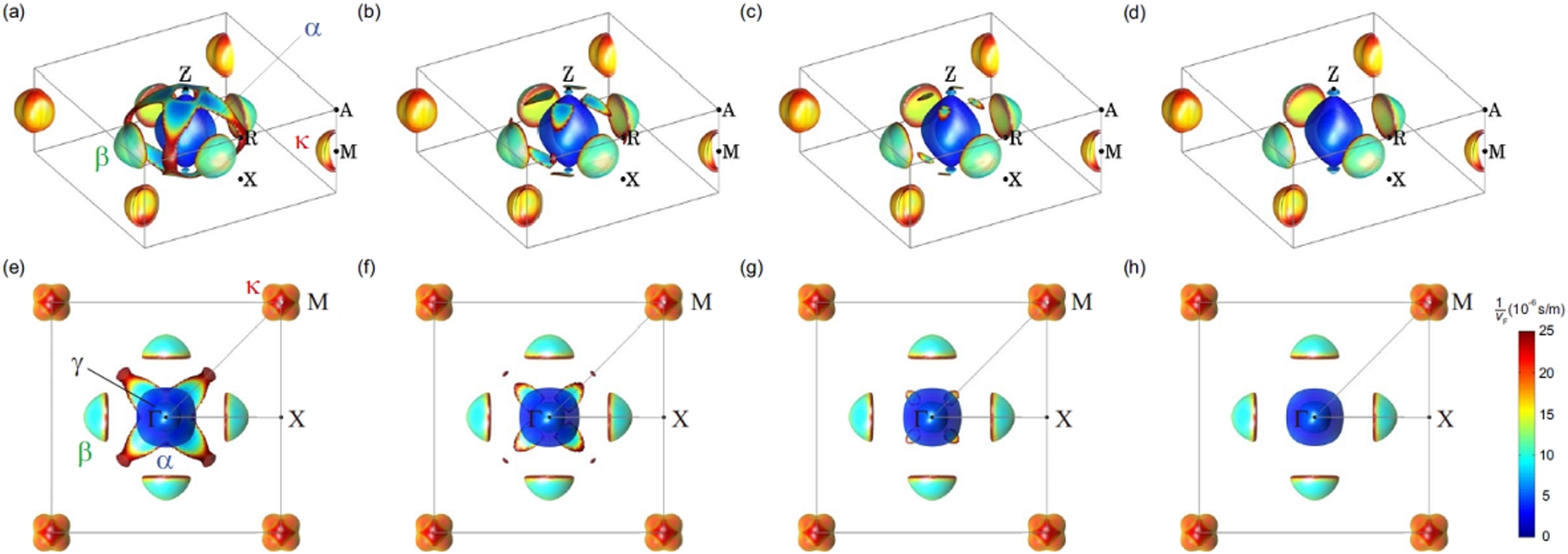}
\caption{(Color online) (a) Schematic views of FS obtained by the density functional band-structure calculations assuming the antiferromagnetic order. The color indicates the inverse of Fermi velocity $1/v_F$ on the FS sheets. The lower is schematic cross sectional view of FS in a plane including $\Gamma$ (the centre), $X$, and $M$ points.  The effective field of 40\,meV is used which corresponds to antiferromagnetic gap of $\sim4$\,meV considering the renormalization of $\sim 1/10$. (b)-(d) Similar calculations for larger effective fields. The energy shifts of 4 (b), 8 (c), and 12\,meV (d) are used for the cage band. We checked that for these small changes of effective field, the main bands $\alpha$, $\beta$ and $\kappa$ do not show any noticeable change.} 
\label{Fig:FS}
\end{figure*} 

The field-angle dependence of the CR lines, which can be compared with the band-structure calculations, allows the determination of the angle-dependent electron masses on the FS sheets in the HO phase. Figures\:\ref{Fig:HO_CR}(a)-(c) display the resonance lines at two different frequencies when the field is inclined from $[100]$ toward $[110]$ direction in the $ab$ plane. It is clear that the CR lines D and A gradually split into two peaks when the field direction is rotated from the $[100]$ to $[110]$ direction. For the line A, the FWHM for finite azimuth angles $\phi$ is significantly broader than that of $\bm{H} \parallel [100]$ $(\phi=0^\circ)$, suggesting that the split occurs immediately after the field rotation from the $a$ axis. On the other hand, the field split occurs at larger angles for the line D. Figure\:\ref{Fig:HO_CR}(f) shows the resonance lines at 60\,GHz when the field is inclined from $[110]$ toward $[001]$ direction. The split in line D survives against the field tilt angle $\theta$ from the basal plane toward the $c$ axis up to the largest tilt $\theta=30^\circ$ used in this study. 

Figures\:\ref{Fig:HO_CR}(d) and (e) display the resonance lines at two different frequencies when the field is inclined from $[100]$ toward $[001]$ direction in the $ac$ plane. The line D exhibits a gradual shift to higher fields with increasing tilt angle $\theta$, whereas the lines B and B' merge into a single peak for large $\theta$. We also note that for the $ac$ plane rotation, due to the limit of the field range we cannot clearly identify the two-peak feature for the line A. Thus we have large error bars for the line A at finite $\theta$ [Fig.\:\ref{Fig:HO_mass}].

\subsection{Assignments of cyclotron mass branches}

The three-dimensional structure of FS mass in the hidden-order phase can be explored by the field-angle dependence of the CR lines, which is summarized in Fig.\:\ref{Fig:HO_mass}. The characteristic angle dependence of each CR line is important to assign the corresponding orbits in different FS pockets. Among the observed CR lines, three lines A (A'), B (B'), and D (D') exhibit strong intensities [solid symbols in Fig.\:\ref{Fig:HO_mass}], which should come from the main FS pockets with relatively large volume. Measurements of the strong lines B and D at two different frequencies provide quantitatively consistent masses, indicating that the mass is field independent at any angle within the measurement range of field. 

The SdH results indicate that the FS in the hidden-order phase is similar to that in the antiferromagnetic state \cite{Has10,Aok12}. Thus we compare our CR results with the band structure calculations assuming the antiferromagnetism \cite{Opp10,Ike12}, and discuss the FS structure in the hidden-order phase. 
The FS structures calculated with several different values of effective field for antiferromagnetism are shown in Fig.\:\ref{Fig:FS}. The results in Fig.\:\ref{Fig:FS}(a) are obtained with the effective field of 40\,meV, which corresponds to an antiferromagnetic gap of $\sim4$\,meV when the renormalization of $\sim1/10$ is taken into account. In this case there exists the cage structure which is absent in the previous calculation \cite{Opp10}, but this structure is sensitive to the gap size. Indeed we found that small energy shifts to this band (4, 8, and 12\,meV for Figs.\:\ref{Fig:FS}(b), (c), and (d), respectively) can diminish this cage structure. Thus our calculations are completely consistent with the previous case with a larger gap. In these calculations in the antiferromagnetic state, we always find main three  non-equivalent FS pockets with relatively large volumes, labelled as $\alpha$, $\beta$, and $\kappa$ [Figs.\:\ref{Fig:FS}(a)-(d)], which have obviously different shapes. Below we show that the $\alpha$, $\beta$, and $\kappa$ bands correspond to the three strong CR lines D (D'), B (B') and A (A'), respectively. The other lines with weaker intensities [open symbols in Fig.\:\ref{Fig:HO_mass}] are likely corresponding to the smaller pockets $\gamma$ inside the $\alpha$ pocket and  hourglass-like small pocket near the $Z$ point [Figs.\:\ref{Fig:FS}(a)-(d) and \ref{Fig:HO_orbit}(a)], as well as the possible remnant pockets of the cage which can be found in some parameter range of antiferromagnetic gap [Figs.\:\ref{Fig:FS}(b) and (c)].  

The $\alpha$ pocket with nearly isotropic shape, which locates around the center of the folded Brillouin zone ($\Gamma$ point), has the largest volume and is the only hole bands among the main bands. According to the magneto-transport measurements, the Hall coefficient is positive in the hidden-order phase \cite{Kas07}, which immediately indicates that this hole band $\alpha$ has much larger mobility than the electron bands in this compensated metal.  This $\alpha$ pocket is therefore responsible for the line D having the strongest intensity and sharpest FWHM (with the largest $\omega_c\tau$). 

There are four $\beta$ electron pockets with hemispherical shape whose center is along the $\Gamma$-$X$ line. These pockets yield two different extremal orbits for $\bm{H}\parallel [100]$, but these two become equivalent for $\bm{H}\parallel [110]$. This uniquely corresponds to the angle dependence of the line B. This line B also tends to merge towards $\bm{H}\parallel [001]$, which is fully consistent with the shape of $\beta$ pockets as well. 

The $\kappa$ pockets around $M$ point have much heavier band mass with larger $1/v_F$ than the $\alpha$ pocket [Figs.\:\ref{Fig:FS}(a)-(d)], which naturally leads us to assign the heaviest line A to the $\kappa$ pockets. Indeed, the $\kappa$ FS consists of two crossing sheets, which should give two different orbits for in-plane fields except when the field is aligned exactly parallel to the $[100]$ direction as observed for line A. As for the $ac$ rotation from $[100]$ to $[001]$ one expects the branch splitting for the $\kappa$ pockets, but the large errors of mass determination for line A at finite $\theta$ prevent us from observing this split clearly. We stress, however, that no splitting in our sharpest line D for the $ac$ rotation toward $[001]$ with much higher resolution is a clear indication that line D does not come from the $\kappa$ band. 

\section{Comparisons with the quantum oscillations}

Our assignments for the $\alpha$ and $\beta$ bands are consistent with the quantum oscillation reports \cite{Ohk99,Has10}, in which the largest amplitude oscillation branch is assigned to $\alpha$ \cite{Ohk99} and the merging branches for $[100]\rightarrow [001]$ to $\beta$ \cite{Has10}. We note that different band assignments to the quantum oscillation branches have been proposed \cite{Opp10}, in which the largest $\Gamma$-centered hole band is assigned to the $\epsilon$ branch observed only at very high fields above $\sim 17$\,T \cite{Shi09}. However, such a high-field branch is most likely associated with field-induced transition \cite{Shi09} possibly due to the Lifshitz topology change by the Zeeman effect \cite{Jo07,Alt11}. It is rather reasonable to assign the most pronounced $\alpha$ branch to this largest hole band, which is consistent with our assignments of CR lines. 

Our cyclotron resonance reveals three strong lines which correspond to the main FS pockets including the heaviest electron band $\kappa$ that has been missing in the quantum oscillation measurements. The heaviest mass and small mean free path of the $\kappa$ pocket as revealed by the large FWHM of line A are likely responsible for the difficulty in observing the corresponding oscillation frequency.

\begin{table*}[t]
 \begin{center}
  \caption{Estimation of the electronic specific heat coefficient $\gamma_i$ for band $i$ from the comparisons with the quantum oscillation (QO) results \cite{Has10}. Values marked with $\dag$ are estimated by assuming the ratio $m^*_{\mathrm{QO}}/m^*_{\mathrm{CR}}=4$.}
   \begin{tabular}{ccccccc} \hline 
band $i$ &$F^{\bm{H}\parallel [100]}$\,(T) & $F^{\bm{H}\parallel [001]}$\,(T)& $m^*_{\mathrm{QO}}(\bm{H}\parallel [001])$&$m^*_{\mathrm{CR}}(\bm{H}\parallel [001])$ &$m^*_{\mathrm{QO}}/m^*_{\mathrm{CR}}$ &$\gamma_i $\,($\frac{\rm mJ}{\mathrm{mol K}^2}$)\\ \hline
$\alpha$ [hole] &1230	&1065	&12.4	&4.4 [D] & 2.8&6.0 \\ 
$\beta(\beta')$ [electron]&219 (751)&422&23.8&6.1 [B] & 3.9&27.3\\ 
$\gamma$ [electron]&73&	195&10	& 2.4 [E] &4.2 &1.4 \\ 
$\kappa$ [electron]&--&-- & (60)$^\dag$ & 15 [A] & (4)$^\dag$ &18.4$^\dag$ \\ \hline
total & & &  & &  &53.1$^\dag$ \\ \hline
   \end{tabular}
   \label{Table:mass}
 \end{center}
\end{table*}

\subsection{Evaluation of the electronic specific heat}

The full determination of the main Fermi surface sheets enables us to evaluate the electronic specific heat coefficient (Sommerfeld constant) $\displaystyle \gamma_{\rm total}=\sum_i \gamma_i$, where $\gamma_i$ is the contribution from band $i$. For closed Fermi surface sheets with spheroidal shape, this can be given by 
\begin{equation}
\gamma_i \approx N_i\frac{k_B^2V}{3\hbar^2} \prod_{j=a,b,c} \left(m_j^* k_F^j\right)^{1/3},
\end{equation} 
where $N_i$ is the number of equivalent sheets within the Brillouin zone for band $i$, $V=49$\,cm$^3$/mol is the molar volume of URu$_2$Si$_2$, $m_j^*$ is the thermodynamic effective mass and $k_F^j$ is the Fermi wave number along $j$ direction. In a spherical band, $k_F$ is directly related to the quantum oscillation frequency $F$ through the extremal cross sectional area $2\pi e F/\hbar = \pi k_F^2$. We approximate each sheet by spheroid with Fermi wave numbers $k_F^j$ $(j=a,b,c)$, which are estimated by using the quantum oscillations results \cite{Has10,Ohk99} for $\bm{H}\parallel [100]$ and $\bm{H}\parallel [001]$. (For $\alpha$ and $\gamma$ sheets we assume $k_F^a=k_F^b$.) The thermodynamic effective mass entered here can be replaced by the mass $m_{\mathrm{QO}}^*$ measured by the quantum oscillations for each band, which is different from $m_{\mathrm{CR}}^*$ determined by the cyclotron resonance \cite{Kan97}. Because the mass $m_{\mathrm{QO}}^*$ has been reported only in a limited field angle range, we simply use $\bm{H}\parallel [001]$ data for each band. We find that the ratio of $m_{\mathrm{QO}}^*$ to $m_{\mathrm{CR}}^*$ is in a range of 3-4 for the main bands of URu$_2$Si$_2$ [Table\,\ref{Table:mass}]. For the heaviest $\kappa$ band, no quantum oscillations have been observed, so we calculate $m_{\mathrm{QO}}^*$ by assuming the ratio $m_{\mathrm{QO}}^*/m_{\mathrm{CR}}^*=4$. 

To evaluate $\gamma_i$ we need the number of sheets $N_i$ for each band, which requires the band assignments. As discussed above, the bands $\alpha$ and $\beta$ are assigned to the hole sheet centered at $\Gamma$ point and the four electron sheets located between the $\Gamma$ and X points [see Figs.\:\ref{Fig:HO_mass} and \ref{Fig:FS}]. The band $\gamma$ with small frequency $F$ and light mass reported by the quantum oscillation experiments \cite{Has10} can be assigned to the small electron pocket inside the $\alpha$ sheet, which we associate with the CR line E with light mass. We ignore the $\eta$ band which has been observed only in a limited range of field angle \cite{Has10}, which likely comes from some extremal orbits on the remnant of the cage with small volume. We also do not consider other branches appeared only at very high fields named as $\delta$ \cite{Jo07}, $\epsilon$ \cite{Shi09}, and $\zeta$ \cite{Alt11}, which may be associated with field-induced transitions. Because of the compensation condition, the volume of $\kappa$ electron bands at the zone corner (which has effectively two pockets) can be estimated by the volumes of one $\alpha$ hole pocket, four $\beta$ electron pockets, and one $\gamma$ electron pocket. From this we can estimate the total Sommerfeld constant $\gamma_{\rm total}$ as large as $\sim53$\,mJ/mol\,K$^2$ [Table\,\ref{Table:mass}], accounting for more than 80\% of the experimental value of $\sim 65$\,mJ/mol\,K$^2$ \cite{Map86}. Considering the assumptions we made, we infer that the agreement is reasonably good. We can make further improvement when we take into account the contributions from the cage with hole-like character, which further make the $\kappa$ volume larger. 

\subsection{Fermi surface topology of the $\alpha$ hole band}

\begin{figure}[t]
\includegraphics[width=0.8\linewidth]{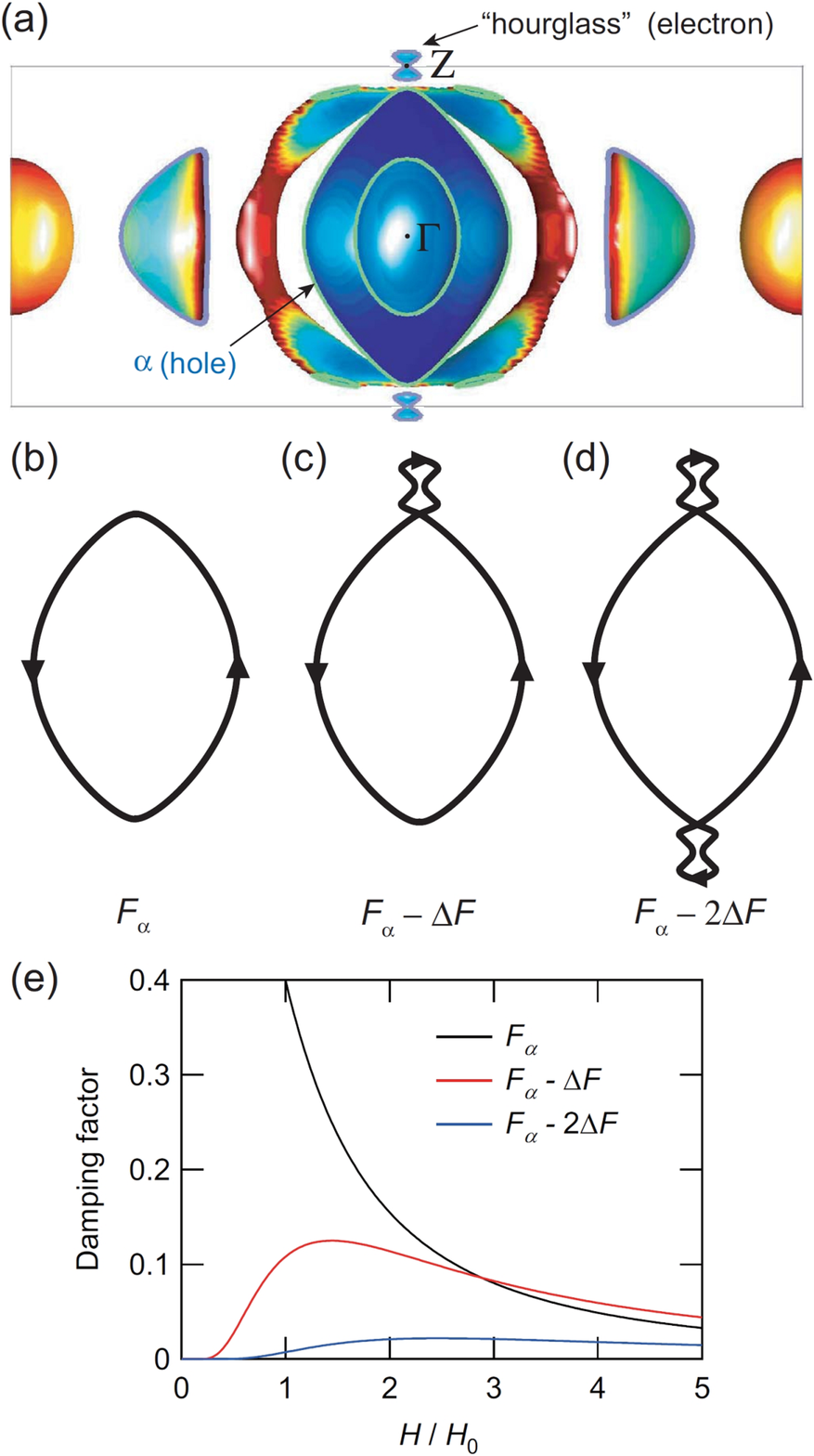}
\caption{(Color online) Possible magnetic breakdown at high magnetic fields for $\bm{H}\perp[001]$. (a) Fermi surface (the same as Fig.\:\ref{Fig:FS}(a)) viewed along the $[100]$ direction. Thin black lines define the Brillouin zone and the colour shades depict the magnitude of $1/v_F$. Thick lines indicate the cross-sectional profile of the FS in a plane including $\Gamma$, Z and X points (see  Fig.\:\ref{Fig:FS}(a)). (b) Cyclotron orbit for the $\alpha$ hole band without magnetic breakdown. Relative probability of performing this orbit is given by $(1-p)^2$, where $p=\exp(-H_0/H)$ is the breakdown probability. (c) `Figure-eight' breakdown orbit through the $\alpha$ hole and `hourglass' electron pockets. 
(d) Orbit with two sets of magnetic breakdown near the two poles of the $\alpha$ band. 
(e) Damping factors for these three orbits which are calculated from the breakdown probability are plotted against field normalized by the breakdown field $H_0$. } \label{Fig:HO_orbit}
\end{figure}

The quantum oscillation experiments show that the oscillation frequency ($F_\alpha$) of the $\alpha$ band changes only weakly with field rotation both within the $ab$ and $ac$ planes, indicating nearly spherical Fermi surface shape \cite{Has10,Ohk99}. 
It is intriguing that for the in-plane fields, the Fourier transfer spectrum of the oscillations above the upper critical field ($\sim 12$\,T for $\bm{H}\perp[001]$) shows multiple four-peak structure in a wide range of azimuth angle $\phi$ \cite{Ohk99,Aok12}. The separation $\Delta F$ between these peak frequencies are nearly angle-independent ($\Delta F\sim 0.07$\,kT), and the number of the oscillation frequencies for $\bm{H}\parallel [110]$ remains the same as that for $\bm{H}\parallel [100]$. This is completely different angle dependence from the branch splitting near the $[110]$ direction found in our cyclotron mass for the line D. 

Very recent Shubnikov-de Haas studies \cite{Aok12} indicate that the three frequencies of the four-peak structure are associated with $\alpha$ band and one lowest frequency comes from the $\beta'$ orbit. We propose that such three frequencies for the $\alpha$ band with a constant separation $\Delta F$ observed only for in-plane fields originates from the field-induced magnetic breakdown effect. Near the $Z$ point of the Brillouin zone, there is a very small electron pocket connected to the next zone [Figs.\:\ref{Fig:FS} and \ref{Fig:HO_orbit}(a)], which has an hourglass shape. The band-structure calculations show that a large electron sheet around the $\Gamma$ point and a slightly smaller hole sheet around the $Z$ point in the paramagnetic state \cite{Opp10,Ohk99} undergo partial gapping by the $\bm{Q}_{C}=(0,0,1)$ zone folding, which results in divided small pockets; the four $\beta$ pockets, the cage, and the `hourglass' pocket \cite{Opp10,Ike12}. This small hourglass pocket is located very close to the $[001]$ pole of the $\alpha$ hole sheet [Figs.\:\ref{Fig:FS} and \ref{Fig:HO_orbit}(a)]. Recent theoretical calculations suggest that several multipole ordered states possible for URu$_2$Si$_2$ have overall similar FS topology as the antiferromagnetic case shown in Figs.\:\ref{Fig:FS} and \ref{Fig:HO_orbit}(a) \cite{Ike12}. 

When a high magnetic field is applied perpendicular to the $[001]$ direction, electrons may tunnel through this small separation between the $\alpha$ hole sheet and the small hourglass electron pocket. The probability of the breakdown occurrence depends exponentially on the field strength, $p=\exp(-H_0/H)$, where the breakdown field $H_0$ depends on the size of the gap. At high enough fields $H\gtrsim H_0$, one can see several different orbits, as shown in Figs.\:\ref{Fig:HO_orbit}(b)-(d), which provides a natural explanation for salient features observed in the quantum oscillation experiments. The breakdown near a $[001]$ pole of the $\alpha$ sheet should decrease the effective oscillation frequency from $F_\alpha$ [Fig.\:\ref{Fig:HO_orbit}(b)] to $F_\alpha-\Delta F$ [Fig.\:\ref{Fig:HO_orbit}(c)]. This is caused by the reduced effective area of the cyclotron orbit with `figure-eight' topology, which is due to a combination of the $\alpha$ hole pocket and the hourglass electron pocket \cite{dHvA}. Another set of breakdown near the other pole leads to the third frequency $F_\alpha-2\Delta F$ [Fig.\:\ref{Fig:HO_orbit}(d)]. Therefore the three-peak structure in the frequency spectrum with the $\phi$-independent separation $\Delta F$ between the peaks can be understood by this mechanism. This also explains the fact that the reduced frequency branches $F_\alpha-\Delta F$ and $F_\alpha-2\Delta F$ disappear once the field direction is inclined from the $ab$ plane, because for such fields the orbits have large separation between the $\alpha$ and hourglass pockets. The breakdown probability of each orbits is $(1-p)^2$ for $F_\alpha$ [Fig.\:\ref{Fig:HO_orbit}(b)], $2p^2(1-p)^2$ for $F_\alpha-\Delta F$ [Fig.\:\ref{Fig:HO_orbit}(c)] and $p^4(1-p)^2$ for $F_\alpha-2\Delta F$ [Fig.\:\ref{Fig:HO_orbit}(d)]. The normalized field dependence of the breakdown probability is shown in Fig.\:\ref{Fig:HO_orbit}(e). From comparisons between these damping factors and quantum oscillation FFT amplitude in the field range from 8 to 15 T \cite{Aok12}, we estimate the breakdown field $H_0$ of the order of $\sim 5$\,T. Figure\:\ref{Fig:HO_orbit}(e) also shows that in the high field range above $\sim 2H_0$, the damping factors of three orbits are of the same order and their ratios have no significant dependence of magnetic field, which also seems to be consistent with the quantum oscillation experiments.

\begin{figure}[t]
\includegraphics[width=0.9\linewidth]{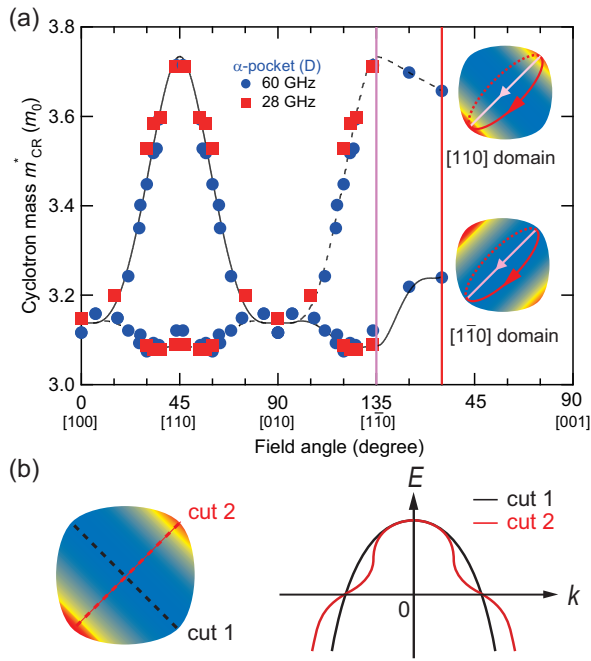}
\caption{(Color online) Structure of the cyclotron mass $m^*_{\rm CR}$ of the $\alpha$ band. (a) $m^*_{\rm CR}$ of the $\alpha$ band as a function of the field angle at 28 and 60 GHz. We use the data for $0 \leq \phi <45^\circ$ and $0 \leq \theta \leq 30^\circ$ and symmetrize them for other angles. The solid and dotted lines are the guides for the eyes representing two domains. Insets illustrate the anisotropic mass distribution in the spherical $\alpha$ band for the two domains elongated along the $[110]$ and $[1\bar{1}0]$ directions. Red area represents heavy spots. For each domain, two orbits for $\phi=-45^\circ$, $\theta=0^\circ$ (pink) and $\phi=-45^\circ$, $\theta=30^\circ$ (red) are depicted. (b) For the $[110]$ domain schematic dispersion curves are shown along the two directions.  }
\label{Fig:HO_mass_alpha}
\end{figure}

The size of the hourglass pocket and hence the magnitude of $\Delta F$ may be sensitive to the details of band-structure calculations, but the present FS in Fig.\:\ref{Fig:HO_orbit}(a) gives $\Delta F\sim 0.03$\,kT, which is the same order as the experimental observations. Recent band-structure calculations suggest that the shape of the $\alpha$ sheet near the poles in the hidden-order phase are sensitive to the order parameters \cite{Ike12}, which may affect the probability of the breakdown at the field range used in the quantum oscillation studies. We stress that our newly found branch splitting of the cyclotron resonance is not originated from this breakdown effect, because our field range ($\mu_0H_{\rm CR}\sim 3$\,T for the $\alpha$ band at 28\,GHz) should be lower than the breakdown field $H_0$. Moreover, the angle dependence of the cyclotron resonance is quite different from the three-peak behavior in the quantum oscillations. These results clearly indicate that the FS shape of the hole $\alpha$ band is nearly spherical, and that the CR mass split results from the peculiar in-plane mass anisotropy which is not directly related to the shape of FS. (In other words, the slope of the energy-momentum dispersion at the $\alpha$ FS is different for different directions while the size of Fermi momentum remains nearly the same [see Fig.\,\ref{Fig:HO_mass_alpha}(b)].) We also note that for the mass determination the cyclotron resonance in a high-quality crystal with large $\omega_c\tau$ ($\sim 20$ in our case) can yield a much higher resolution than the quantum oscillation measurements, which require the analysis of temperature dependent oscillation amplitude. This allows us to expose the anisotropic mass structure which has been hidden in the hidden-order phase of URu$_2$Si$_2$.

\section{Nematic electronic structure inferred from cyclotron resonance}

\subsection{In-plane mass anisotropy of the $\alpha$ hole band}
Now we focus on the signature of the FS that provides a key to understanding the HO.  The most unexpected and important result is that the sharpest line D arising from the $\alpha$ hole pocket is clearly split into two lines with nearly equal intensities near the $[110]$ direction [Figs.\:\ref{Fig:HO_CR}(a), (c), (f) and \ref{Fig:HO_mass}]. This splitting in the $\alpha$ hole pocket is hardly explained from the calculated FS structure in the antiferromagnetic phase. 

One may argue that some small warping of FS shape gives rise to the appearance of an additional extremal orbit for particular angles, which can result in the splitting. However, such a scenario is highly unlikely because of the following reasons.  First, quantum oscillation measurements clearly indicate that the number of extremal orbits in the $[110]$ direction remains the same as that in $[100]$ direction \cite{Ohk99,Aok12}. This argues against that the additional orbit appears only near the $[110]$ direction.  Second, the band-structure calculation indicates that $\alpha$ pocket is nearly spherical, which is supported by the quantum oscillations experiments \cite{Ohk99,Has10} that reveal almost angle-independent oscillation frequency for this band.  Third, the fact that the integrated intensity of the split line D' is nearly equal to that of D line in a wide range of angle [Figs.\:\ref{Fig:HO_CR}(a), (c) and (f)] suggests that both CR lines arise from the orbits with nearly equal FS cross sections, which is at odds with the warping scenario. 
Therefore we infer that the observed splitting of the CR line D results from a peculiar mass structure in the HO phase. 

The observed splitting of the sharpest CR line D near the $[110]$ direction can be naturally explained by the emergence of the heavy spots in the $\alpha$ band along $[110]$ as depicted in Fig.\:\ref{Fig:HO_mass_alpha}. One may consider such a twofold mass anisotropy if the system breaks the fourfold symmetry as suggested by the magnetic torque experiments \cite{Oka11}. The twofold symmetry also leads to the formation of domains with different nematic directions, which can be called as $[110]$ and $[1\bar{1}0]$ domains as shown in the insets of Fig.\:\ref{Fig:HO_mass_alpha}(a). In large crystals containing these two domains we expect to have two different orbits with different cyclotron masses near the $[110]$ and equivalent directions. This can explain the observed splitting of the cyclotron mass near $[110]$. The robustness against field tilting can also be explained because the cyclotron orbit for one of the domains always goes through the heavy spots as shown by the red arrow in the inset of Fig.\:\ref{Fig:HO_mass_alpha}(a). 

The fact that the quantum oscillation frequency for the $\alpha$ band does not show the corresponding splitting for tilted fields indicates that the FS shape does not show significant breaking of fourfold symmetry but only mass structure strongly breaks the rotational symmetry. This suggests that the twofold symmetry is related to the correlation effect, which modifies mainly the curvature of band dispersion near the Fermi energy [Fig.\:\ref{Fig:HO_mass_alpha}(b)].

\subsection{Enhanced inelastic scattering rate at the heavy spots}

\begin{figure}[b]
\includegraphics[width=\linewidth]{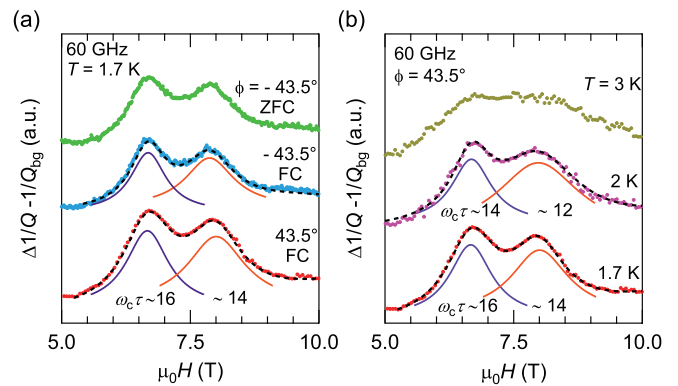}
\caption{(Color online) Split CR lines for the $\alpha$ band near the [110] direction. (a) Field dependence of $\Delta 1/Q-1/Q_{\rm bg}$ near the resonance line D at 60\,GHz for $\phi=\pm 43.5^\circ$. The data taken after the zero-field cooling (ZFC) and field-cooling (FC) procedures are identical within experimental error. The FC data are fitted by the two Lorentzian functions (dashed lines) with different $\omega_c\tau$ values (solid lines). Each curve is shifted vertically for clarity. (b) The same plot for $\phi=43.5^\circ$ at three different temperatures. }
\label{Fig:HO_split}
\end{figure}

\begin{figure}[t]
\includegraphics[width=0.9\linewidth]{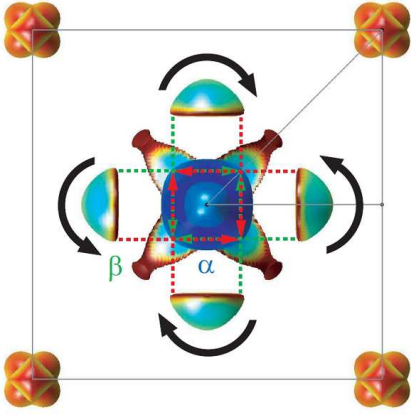}
\caption{(Color online) Schematic Fermi surface viewed along the $[001]$ direction, showing possible changes due to the fourfold rotational symmetry breaking. In a $[110]$ nematic state, the four hemispherical $\beta$ pockets may rotate or elongate towards each directions indicated by the black arrows. This symmetry breaking changes the nesting conditions between the $\beta$ and $\alpha$ pockets connected with the incommensurate wave vectors $(\pm0.4,0,0)$ and $(0,\pm0.4,0)$ (dashed arrows), and the interband quasiparticle scattering rates shown by the green and red arrows become non-equivalent. }
\label{Fig:HO_spot}
\end{figure}
A close look at the split resonance lines in Fig.\:\ref{Fig:HO_split} reveals that the FWHM for the heavier line (at higher field) is always larger than the lighter line in our measurement temperature range. We note that the integrated intensities for these two CR lines are almost identical, which is reproducible when the field direction is rotated by $\sim 90^\circ$. This implies that the two domains have almost identical volumes in the large crystal used in this study. We also find that the data taken after the field cooling condition at 12\,T remains unchanged, which is consistent with the previous torque experiments \cite{Oka11}. These results suggest that the domains are pinned by the underling lattice conditions. 

An important result is that the magnitude of $\omega\tau$ estimated from the width of the Lorentzian fits is smaller for the heavier resonance line. It has been demonstrated \cite{Ebi92} that the elastic impurity scattering time is longer for heavier band, because of the impurity limited mean free path at very low temperatures. In our case, however, the opposite trend has been observed, implying that the inelastic scattering rate $1/\tau$ is enhanced for the heavier cyclotron orbit. This is supported by the higher temperature data [Fig.\:\ref{Fig:HO_split}(b)], where the FWHM of the heavier CR line shows more broadening, indicating stronger temperature dependence of inelastic scattering rate $1/\tau(T)$ near the heavy spots.

To discuss possible origins of the enhanced inelastic scattering at the heavy spots, it is important to consider the interband scattering between the Fermi surface points connected with particular wave vectors characteristic to the hidden-order phase. It has been shown from the neutron scattering experiments \cite{Wie07,Vil08} that the excitations at the commensurate $\bm{Q}_{C}=(1,0,0)=(0,0,1)$ and incommensurate $\bm{Q}_{IC}=(0.4,0,0)$ wave vectors are important in the hidden-order phase of URu$_2$Si$_2$. As schematically shown in Fig.\:\ref{Fig:HO_spot}, there are parts of $\alpha$ and $\beta$ pockets connected approximately by $\bm{Q}_{IC}$ \cite{Opp10}. In the HO phase, the FS has twofold symmetry (rather than fourfold tetragonal symmetry) along the $[110]$ direction, and the four $\beta$ pockets can be slightly rotated (or elongated) along this direction. This leads to an imbalance in the interband interactions as depicted by the green and red dashed arrows in Fig.\:\ref{Fig:HO_spot}. This imbalance may enhance the inelastic scattering at the two of the four corners of the $\alpha$ hole pocket, generating the hot spots along the $[110]$ (or $[\bar{1}10]$) direction in the $\alpha$ sheet as revealed in the present CR experiments.

\begin{figure}[b]
\includegraphics[width=1.0\linewidth]{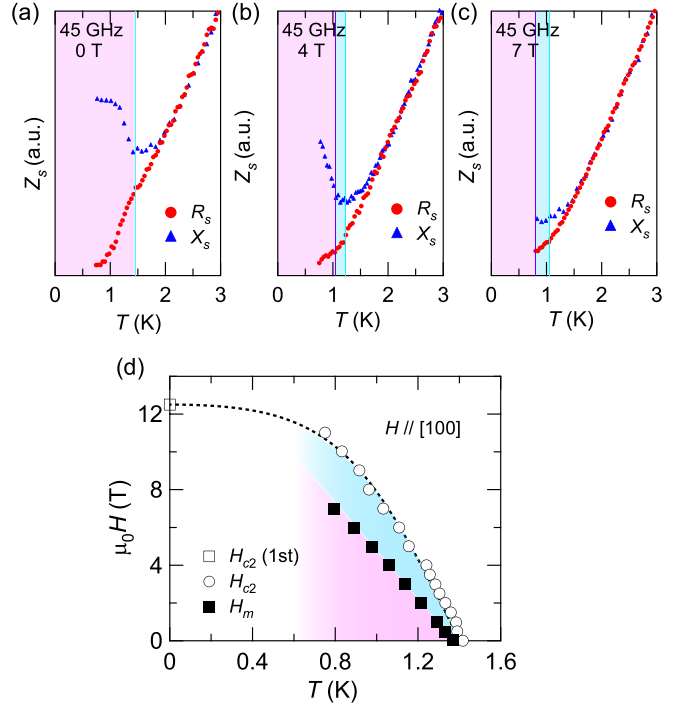}
\caption{(Color online) (a) Temperature dependence of surface resistance $R_s$ and surface reactance $X_s$ at 45 GHz in URu$_2$Si$_2$ at zero field.  Shaded region represents the superconducting state below $T_{\rm SC}$. (b), (c) Similar plots for data taken under magnetic fields $\bm{H} \parallel [100]$. (d) Field-temperature phase diagram for $\bm{H} \parallel [100]$ taken from Ref.\:\onlinecite{Oka08}. In (a)-(c), the corresponding temperature regions for the vortex liquid (blue shade) and vortex solid (red shade) states are also indicated.}
\label{Fig:SC}
\end{figure}

\section{Cyclotron resonance in the superconducting state}

\subsection{Temperature dependence of surface impedance}

Next we discuss the results in the superconducting state, which have been obtained by using 45-GHz cavity with $^3$He pot. To check the superconducting transition we measure the temperature dependence of the surface impedance ($Z_s = R_s + {\rm i} X_s$) at zero and finite magnetic fields.  In the condition that the skin depth $\delta$ is smaller than sample dimensions, the change in the microwave power dissipation $\Delta1/Q$ and the frequency shift $\Delta f$ are proportional to the surface resistance $R_s$ and the change of the surface reactance $\Delta X_s$, respectively \cite{Shibauchi94}. In many superconductors, the normal-state microwave electrodynamics is described by the Hagen-Rubens limit $\omega \tau \ll 1$, in which the surface resistance and reactance have the simple form 
\begin{equation}
R_s = X_s = \sqrt{\frac{\mu _0 \omega \rho }{2}} = \frac{\mu _0 \omega \delta }{2}.
\label{eq_HR_limit}
\end{equation} 
However, in the present case $\omega \tau$ becomes large and the Hagen-Rubens limit is not satisfied at low temperatures. Indeed the temperature dependence of  $X_s$ deviates from that of $R_s$ even above $T_{\rm SC}$, as shown in Fig.\:\ref{Fig:SC}(a). Similar deviations have been observed, for example, in the Kondo semiconductor CeNiSn, where the gap formation reduces the scattering leading to $\omega \tau>1$ at low temperatures\cite{Shibauchi97}.

\begin{figure}[b]
\includegraphics[width=0.75\linewidth]{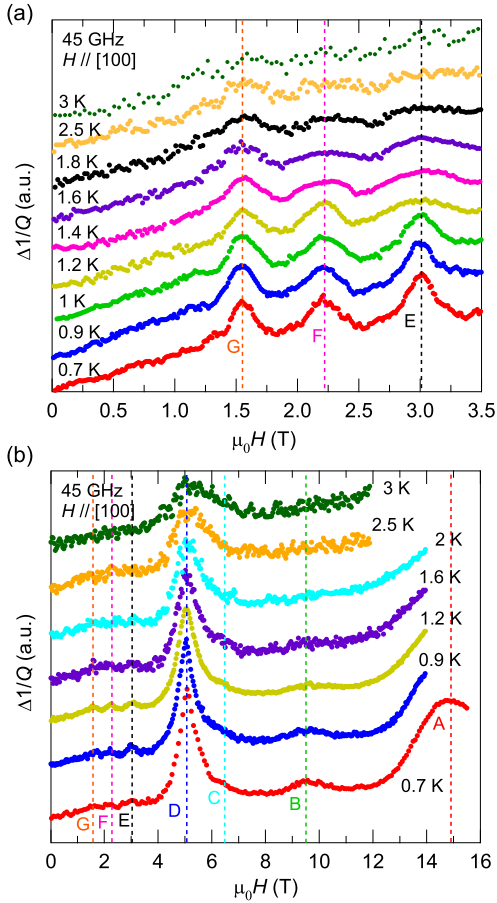}
\caption{(Color online) Observation of cyclotron resonance in the superconducting phase of URu$_2$Si$_2$.  (a) Field dependence of the change $1/Q$ at several temperatures for $\bm{H}\parallel [100]$.  }
\label{Fig:SC_CR}
\end{figure}

In the SC state below $T_{\rm SC}$, we clearly observe the reduction of $R_s$ as well as the enhancement of $X_s$ [Fig.\:\ref{Fig:SC}(a)]. The latter enhancement is not usually seen in superconductors, because $X_s$ in the SC state can be approximately given by the magnetic penetration depth $\lambda $ as
\begin{equation}
X_s = \mu _0 \omega \lambda,
\end{equation} 
when $\lambda$ is much shorter than the skin depth $\delta$.
In URu$_2$Si$_2$, however, the small number and heavy mass of carriers result in very long penetration depth $\sim 1\,\mu$m \cite{Ama97}, which is longer than the skin depth $\sim 0.24\,\mu$m at 45\,GHz in our ultraclean crystals. 
Under magnetic fields, we observe essentially similar behaviours in the temperature dependence of $Z_s$ [Figs.\:\ref{Fig:SC}(b) and (c)]. The deviation between $R_s(T)$ and $X_s(T)$ is observed in the normal state, which is enhanced in the SC state. Previous measurements of electrical and thermal conductivities in fields \cite{Oka08} have provided evidence for the existence of extended region of vortex liquid state, where the resistivity is still finite even in the SC state below the mean-field transition temperature $T_{\rm SC}$.  In the vortex solid state below the melting temperature $T_m$, the resistivity becomes zero and the thermal conductivity shows an enhancement \cite{Oka08}. By a comparison with the reported vortex phase diagram [Fig.\:\ref{Fig:SC}(d)], we find that the deviation between $R_s(T)$ and $X_s(T)$ becomes much enhanced below $T_m$ in the vortex solid state.

\begin{figure}[b]
\includegraphics[width=0.8\linewidth]{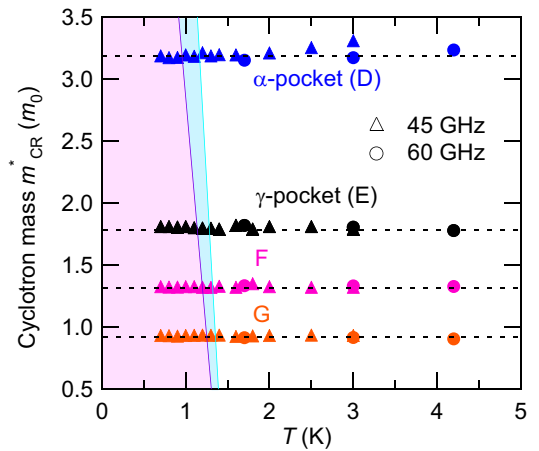}
\caption{(Color online) The temperature dependence of the cyclotron masses at 45  and 60\,GHz for CR lines D-G. Corresponding vortex liquid (blue shade) and vortex solid (red shade) states for 45\,GHz are determined from the phase diagram in Ref.\:\onlinecite{Oka08}.}
\label{Fig_SC_mass}
\end{figure}

\subsection{Cyclotron mass in the superconducting state}

The field dependence of $\Delta 1/Q$ for $\bm{H}\parallel [100]$ at 45\,GHz shows multiple CR lines, which persist down to the lowest temperature at 0.7\,K [Figs.\:\ref{Fig:SC_CR}(a) and (b)]. At this temperature 0.7\,K, the upper critical field $H_{c2}$ and vortex-lattice melting field $H_m$ for $\bm{H}\parallel [100]$ are 11 and 7\,T respectively \cite{Oka08}. Thus we focus on the resonance lines at low fields, labelled as D, E, F, and G. The surface impedance results in Fig.\:\ref{Fig:SC} clearly indicate that we cover the field and temperature range deep in the SC state, and thus for these lines the observed clear peaks in the SC state below $T_{\rm SC}$ can be considered as the first observation of the CR in the SC phase of heavy-fermion materials.

The temperature dependence of the cyclotron masses $m^*_{\rm CR}$ for these lines, which are determined by the resonance fields, is plotted in Fig.\:\ref{Fig_SC_mass}. Within experimental error, the masses are temperature independent, and we find that they do not exhibit any noticeable change between the HO and SC phases. Theoretically, in clean type-II superconductors, it has been shown \cite{Kopnin2001} that the presence of superconducting and normal carriers in the vortex states leads to the violation of Kohn's theorem, and the temperature dependence of the superconducting carrier number results in a peculiar temperature dependence of the cyclotron frequency. In the present case, however, such temperature dependence is not observed. One possible reason for this difference is that in this multiband system, the Kohn's theorem is already violated in the normal state above $T_{\rm SC}$, which may alter the description of CR in the vortex state. Further studies are necessary for the understanding of CR in heavy-fermion and multiband superconductors.

\subsection{Temperature dependence of the scattering rate}

\begin{figure}[t]
\includegraphics[width=1.0\linewidth]{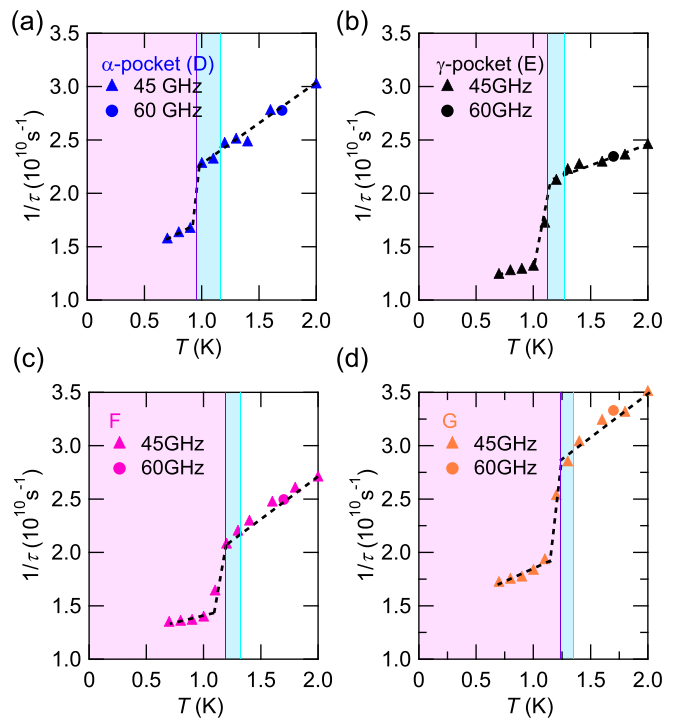}
\caption{(Color online) Temperature dependence of the scattering rate $1/\tau$ extracted from the CR line width at 45 and 60\,GHz for CR lines D (a), E (b), F (c), and G (d).  Corresponding vortex liquid (blue shade) and vortex solid (red shade) states for 45\,GHz are determined from the phase diagram in Ref.\:\onlinecite{Oka08}.}
\label{Fig_SC_tau}
\end{figure}

A close look at the temperature dependence of the resonance line shape in Figs.\:\ref{Fig:SC_CR}(a) and (b) finds that the FWHM of the CR lines become broader with increasing temperature. We analyze the data by the Lorentzian fits to extract the temperature dependence of the quasiparticle scattering rate $1/\tau$, which is shown in Fig.\:\ref{Fig_SC_tau}. In the normal state above $T_{\rm SC}$ the scattering rate roughly follows $T$-linear dependence, which suggests the deviation from the standard Fermi-liquid theory of metals. Such non-$T^2$ dependence of scattering can also be found in the transport measurements of URu$_2$Si$_2$ \cite{Mat11,Zhu09}. 

In the SC state below $T_{\mathrm{SC}}$, we observe that the scattering rate is suppressed not at $T_{\mathrm{SC}}$, but below the vortex-lattice melting transition temperature $T_m$ for all four lines D-G [Figs.\:\ref{Fig_SC_tau}(a)-(d)]. 
Such a suppression below $T_m$ is consistent with the reported enhancement of thermal conductivity below $T_m$, and provides further evidence that the vortex solid state has much less scattering than the vortex liquid state. In very clean systems the vortices form a regular lattice in the solid state, which can act as a periodic potential for quasiparticles. In such a case, the formation of a Bloch-like state may be expected, where the quasiparticle scattering is reduced compared with the vortex liquid state having more disordered potentials. Thus our observation provides a strong support for the formation of the quasiparticle Bloch state in vortex-lattice state of clean superconductors. 

We also note that the dHvA measurements \cite{Ohk99} have shown that the Dingle temperature $T_D$, which is proportional to the scattering rate, increases when the field is reduced below the upper critical field at very low temperatures $\sim 35$\,mK. In this low temperature range, the upper critical field is suggested to be of first order \cite{Kas07}, and the boundary between the vortex liquid and solid states is not fully resolved. This result can be explained by the enhanced scattering due to the Andreev reflection occurred at the boundary between the SC and normal (vortex core) regions in the vortex liquid state just below $H_{c2}$. This implies that there is sizeable vortex liquid region even in the low-temperature limit, where thermal fluctuations vanish. Such a possible quantum vortex liquid state, where quantum fluctuations melt the vortex lattice, has been suggested in organic and cuprate superconductors \cite{Sasaki1998, Shibauchi2003}. It deserves further studies to determine the complete phase diagram down to the low temperature limit.

\section{Concluding remarks}

We have shown that the observation of CR in the HO and SC states of URu$_2$Si$_2$ provides useful information of the quasiparticle mass and scattering rate. 

In the HO phase, we have determined the angle-dependent mass structure, which is compared with the band-structure calculations assuming the antiferromagnetic state. We were able to assign the three CR lines with strong intensities with the main Fermi surface pockets, from which a reasonable estimate of electronic specific heat coefficient can be obtained. In the sharpest CR line corresponding to the hole $\alpha$ pocket, we observed anomalous splitting near the $[110]$ direction, which is not expected in the calculations for antiferromagnetic state. This splitting is found to be robust against the field tilting from the basal plane, which clearly indicates that the CR splitting has a different origin from the threefold splitting of the quantum oscillation frequency observed only near the in-plane direction. By considering the Fermi-surface structure we propose the magnetic breakdown at high fields as a possible origin of the quantum oscillation splitting. The CR splitting can be naturally explained if we consider the twofold in-plane anisotropy of the mass and the formation of micro-domains, which is consistent with the broken rotational fourfold symmetry suggested by the magnetic torque experiments \cite{Oka11}. Recent nuclear magnetic resonance (NMR) experiments have reported the peculiar in-plain anisotropy of the Si NMR line width, which also supports the twofold anisotropy elongated along the $[110]$ direction \cite{Kam13}. Moreover, very recent high-resolution synchrotron X-ray measurements have revealed evidence for structural change from the tetragonal $I4/mmm$ to orthorhombic $Fmmm$-type symmetry \cite{Ton13}, which is also consistent with the broken fourfold symmetry. From the scattering rate analysis, we find that the anisotropic in-plane mass structure involves the hot spots with heavy mass and large scattering rate, which suggests strong momentum dependence of electron correlations due to interband scattering. 

The broken fourfold symmetry gives strong constraints on the symmetry of the order parameter in the HO phase. In the symmetry classification of the multipole order \cite{Kis05,Tha11,Suzuki}, this is consistent with the two-dimensional $E$ representations. Along this line, rank-2 quadrupole with $E^+$ symmetry \cite{Tha11}, and rank-3 octupole \cite{Han12} and rank-5 dotriacontapole with $E^-$ symmetry \cite{Ike12,Rau12} have been theoretically proposed, where the superscript $+$ or $-$ denotes the parity with respect to time reversal. More exotic nematic or hastatic states with and without time reversal symmetry breaking have also been proposed \cite{Fuj11,Ris12,Cha13}, which are also consistent with the broken fourfold symmetry. To pin down the genuine HO parameter, the next important step would be to identify whether time reversal symmetry is broken or not \cite{Shi12}. Recent NMR analysis suggests a time reversal broken state \cite{Tak11}, and more experiments are welcome to establish this issue.  

In the SC phase embedded in the HO phase, we presented the first observation of CR in the vortex states. We found that while the cyclotron mass does not show any temperature dependence, the scattering rate shows a rapid suppression below the vortex lattice melting transition temperature. This provides evidence for the formation of quasiparticle Bloch state in the vortex lattice state.  

\section*{acknowledgments}
We thank D. Aoki, K. Behnia, 
P. Chandra, P. Coleman, R. Flint, S. Fujimoto, 
G. Knebel, H. Kontani, G. Kotliar, Y. Kuramoto, 
J.\,A. Mydosh, K. Miyake, 
M.-T. Suzuki, T. Takimoto, P. Thalmeier, C.\,M. Varma, 
and H. Yamagami for helpful discussions. This work was supported by 
Grant-in-Aid for the Global COE program ``The Next Generation of Physics, Spun from  Universality and Emergence'', Grant-in-Aid for Scientific Research on Innovative Areas ``Heavy Electrons'' (No.\,20102002, 20102006, 23102713) from MEXT, and KAKENHI from JSPS. 

\end{document}